\documentclass[conference]{IEEEtran}
\IEEEoverridecommandlockouts
\PassOptionsToPackage{hyphens}{url}
\usepackage{hyperref}
\usepackage{cite}
\usepackage{amsmath,amssymb,amsfonts}
\usepackage{algorithmic}
\usepackage{graphicx}
\usepackage{textcomp}
\usepackage{xcolor}
\def\BibTeX{{\rm B\kern-.05em{\sc i\kern-.025em b}\kern-.08em
    T\kern-.1667em\lower.7ex\hbox{E}\kern-.125emX}}
\usepackage{listings}
\usepackage{tikz}

\usepackage{comment}

\definecolor{babyblueeyes}{rgb}{0.63, 0.79, 0.95}
\definecolor{antiquewhite}{rgb}{0.98, 0.92, 0.84}

\begin{document}

\title{Neural Fault Injection: Generating Software Faults from Natural Language}

\author{\IEEEauthorblockN{Domenico Cotroneo}
\IEEEauthorblockA{
\textit{University of Naples Federico II}\\
Naples, Italy \\
cotroneo@unina.it}
\and
\IEEEauthorblockN{Pietro Liguori}
\IEEEauthorblockA{
\textit{University of Naples Federico II}\\
Naples, Italy \\
pietro.liguori@unina.it}}

\maketitle
\thispagestyle{plain}
\pagestyle{plain}

\begin{abstract}
Traditional software fault injection methods, while foundational, face limitations in adequately representing real-world faults, offering customization, and requiring significant manual effort and expertise. This paper introduces a novel methodology that harnesses the capabilities of Large Language Models (LLMs) augmented with Reinforcement Learning from Human Feedback (RLHF) to overcome these challenges. 
The usage of RLHF emphasizes an iterative refinement process, allowing testers to provide feedback on generated faults, which is then used to enhance the LLM's fault generation capabilities, ensuring the generation of fault scenarios that closely mirror actual operational risks. This innovative methodology aims to significantly reduce the manual effort involved in crafting fault scenarios as it allows testers to focus on higher-level testing strategies, hence paving the way to new possibilities for enhancing the dependability of software systems. 
\end{abstract}

\begin{IEEEkeywords}
Software Fault Injection, Large Language Models, Reinforcement Learning from Human Feedback, Natural Language Processing
\end{IEEEkeywords}

\section{Introduction}
\label{sec:introduction}
Ensuring the dependability of software systems remains one of the most pressing challenges in the realm of computer science~\cite{ferreira2021reliability,cotroneo2020dependability,cai2021s}. As these systems grow in complexity and scale, traditional methodologies for identifying and mitigating potential faults increasingly struggle to keep pace~\cite{kasauli2021requirements,maciel2021survey}. 

Software fault injection has historically provided valuable means to ensure that the system will be safe even in the presence of residual software faults (or bugs)~\cite{natella2016assessing,madeira2000emulation}. However, existing approaches to fault injection are often limited by the heavy reliance on predefined fault models, which may not adequately represent realistic failure scenarios encountered in real-world applications~\cite{costa2015practical,natella2012fault}. 
Furthermore, significant technical expertise and manual effort for configuration and execution are required for these approaches to be implemented successfully\cite{huang2012taxonomy,huang2017human}.
As a result, it is really difficult to achieve representativeness, and thus effectiveness, at an acceptable cost in real-world application scenarios. 
Therefore, while software fault injection remains a cornerstone of dependability testing, the limitations of existing tools and methodologies pose significant challenges.
Addressing these challenges requires a fundamental rethinking of the approach to software fault injection, paving the way for more innovative, flexible, and efficient solutions.

In light of these considerations, we envision a new research vision for software fault injection, where a new generation of approaches will be able to bridge the gap between complex and realistic fault scenarios and their practical and effective implementation in testing environments.  
For example, we refer to a situation where a tester can describe a fault scenario in natural language (NL), like  ``\textit{introduce a race condition between processes A and B when condition C is met}", and the system will produce the corresponding faulty code to simulate this condition. This results in more reliable and robust software systems by simplifying the testing process and broadening the range of possible fault scenarios that may be tested.

To achieve this goal, this paper presents a forward-looking and disruptive solution to the challenge of ensuring software dependability by proposing an innovative methodology to perform software fault injection. The methodology leverages the capabilities of Large Language Models (LLMs), a class of advanced artificial intelligence systems designed to understand, generate, and interact with human language at scale, augmented with Reinforcement Learning from Human Feedback (RLHF), i.e., a method to align language models with user descriptions by fine-tuning them with human feedback~\cite{ouyang2022training}. This combination creates a dynamic, automated process that improves the generation of simulated software faults, tailored to testers' specific needs. 

More precisely, the adoption of LLMs, which have already shown their capabilities in several software engineering tasks~\cite{hou2023large,fan2023large}, enables the
translation of natural language descriptions of fault scenarios directly into executable code for software testing, hence reducing the manual effort required to design and implement fault scenarios. 
In this process, RLHF plays a pivotal role by refining the alignment of LLMs with user-specific descriptions and employing a fine-tuning process that incorporates human feedback directly into the model's learning cycle. 

We describe the proposed methodology by detailing an end-to-end process from the initial fault definition using natural language to the automated integration of generated faults into the software's codebase for testing. Moreover, we discuss the practical implementation of this solution and its feasibility across various stages of software development and testing. 

In the following,
Section~\ref{sec:problem} illustrates the problem statement; Section~\ref{sec:methodology} describes the proposed methodology; Section~\ref{sec:threats} discusses the threats to validity;
Section~\ref{sec:conclusion} concludes the paper.

\section{Problem Statement}
\label{sec:problem}
As software systems evolve, becoming more complex and integrated into critical aspects of society, the stakes for ensuring their reliability, safety, and security have never been higher. Software fault injection (SFI) has emerged as a pivotal technique in this endeavor~\cite{cotroneo2019bad,cotroneo2020fault}, aiming to assess and enhance the reliability of systems by deliberately introducing faults and observing their behavior. Despite their critical role, the current landscape of SFI tools and methodologies exhibits significant limitations that hamper their effectiveness and applicability to modern software development paradigms. In the following, we provide an overview of these limitations.

\subsubsection{Inadequate Representation of Real-World Faults}
One of the primary limitations of existing SFI tools is their reliance on a limited set of predefined fault models~\cite{qasystemscantata,razorcattessy}. These models often fail to describe the full spectrum of faults that can occur in real-world scenarios, particularly those involving complex interactions among software components or emerging technologies such as cloud computing and microservices. 
Current tools might model simple faults like function failures but fall short in simulating complex scenarios such as race conditions in concurrent processing, or the complexity of cloud service outages affecting distributed systems, where network latency and transient failures are common. This gap between theoretical fault models and practical fault scenarios undermines the comprehensiveness of testing, leaving systems potentially vulnerable to unanticipated failures.

\subsubsection{Lack of Customization and Flexibility}
Traditional SFI tools offer limited scope for customization~\cite{cotroneo2013fault,van2016hsfi,schwahn2018fastfi}, forcing testers to adapt their testing strategies to the constraints of the tools rather than the specific needs of their software systems. This one-size-fits-all approach is increasingly misaligned with the diverse and specialized requirements of contemporary software applications, which demand more nuanced and flexible testing methodologies that can be tailored to the unique characteristics of each system. 

\subsubsection{Manual Effort and Expertise Requirements}
The effective use of conventional SFI tools often requires a significant investment in manual effort and specialized expertise~\cite{winter2011impact,giuffrida2013edfi}. Testers must not only select appropriate fault models but also configure them correctly for each testing scenario. This process is time-consuming and prone to human error, making it a bottleneck in the testing cycle. Moreover, the need for deep expertise in both the tools themselves and the underlying fault models can limit accessibility for testers, especially in smaller teams or less resource-rich environments.

Addressing these challenges requires a fundamental rethinking of the approach to software fault injection, paving the way for more innovative, flexible, and efficient solutions. Our proposed methodology seeks to bridge this gap by leveraging the adaptability and learning capabilities of LLMs to generate realistic, complex fault scenarios tailored to the specific context of the application under test and, at the same time, reduce the manual effort and expertise required from testers.

\section{Proposed Methodology}
\label{sec:methodology}
\begin{figure}[t]
\centering
\includegraphics[width=1\linewidth]{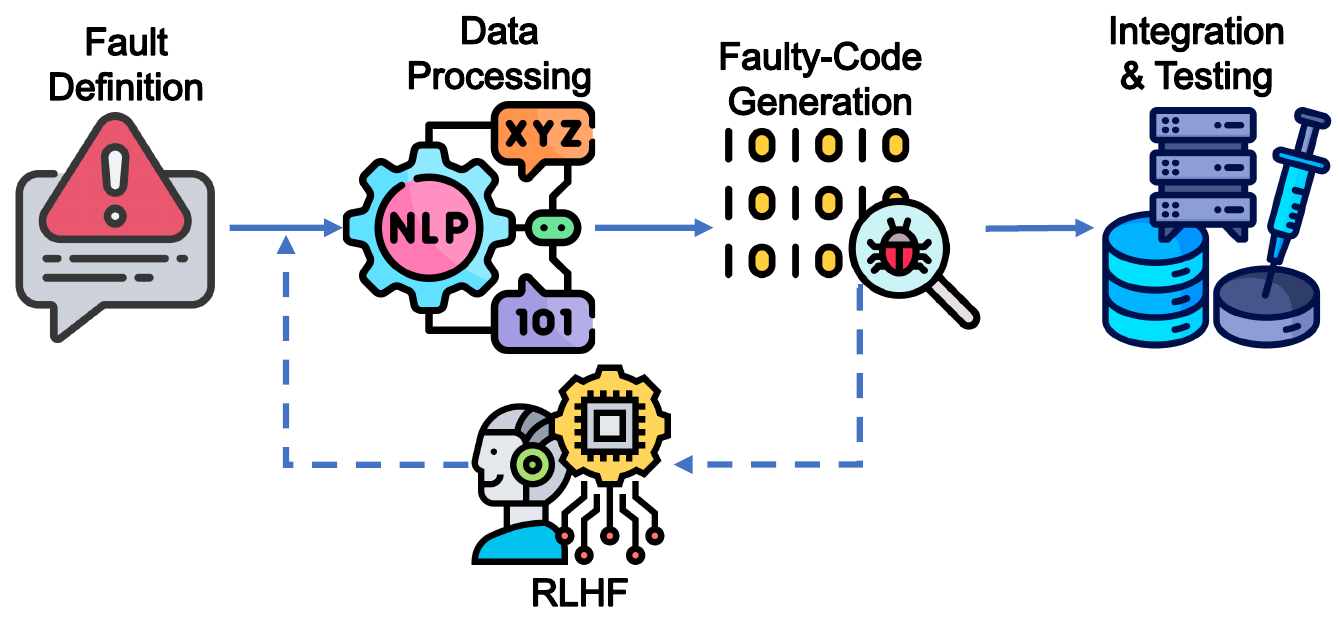}
  \caption{Workflow of the proposed methodology.}
  \label{fig:methodology}
\end{figure}

We propose a novel methodology for software fault injection that leverages the capabilities of LLMs enhanced by RLHF, a machine-learning technique that uses human feedback to optimize models to self-learn more efficiently. This process is designed to automate and refine the generation of simulated software faults based on testers' inputs, making the testing of software systems both more efficient and effective.  \figurename{}~\ref{fig:methodology} illustrates the workflow of the methodology, which we detail in the following.

The initial step involves fault definition, a pivotal step where testers articulate potential fault scenarios in natural language, complementing this with the submission of relevant code segments or complete files. The use of natural language is key to capturing complex, real-world fault scenarios that might not be readily identified through traditional testing methods. This approach not only facilitates the generation of simulations that are both technically accurate and highly relevant to the testers' goals but also specifically targets the uncovering and simulation of residual faults.

Residual faults, which are those subtle and often overlooked errors that remain in the system after initial deployment, pose a significant challenge for traditional fault injection methods. Our dual-input approach allows for the exploration of complex interactions and edge cases within the software that might harbor these elusive errors. The methodology’s strength lies in its ability to simulate not just the faults testers can predict or have previously encountered, but also those that are unforeseen, thus significantly enhancing the software's reliability.

Following the fault definition, the natural language input undergoes data processing handled by a Natural Language Processing (NLP) engine. This phase involves dissecting the tester's description and restructuring it into a format that is tailored for LLM interpretation. This step ensures the LLM's comprehensive understanding of the fault scenario, converting intuitive descriptions into a structured format that aligns with the operational framework of the LLM. Achieving this understanding is pivotal, as it ensures the generated fault simulation is precisely aligned with both the technical and contextual requirements of the specified software environment.

The core of our methodology lies in the code generation phase, where the LLM, leveraging its extensive training across diverse codebases and fault patterns, crafts a code snippet that embodies the specified fault. This generated code is a testament to the LLM's ability to produce realistic and relevant faulty code snippets that address the tester's inputs. This process is further enhanced by the integration of RLHF. Indeed, testers review the generated code, providing targeted feedback that is immediately harnessed to refine the LLM's outputs. This feedback loop is instrumental in evolving the LLM's fault generation capabilities, ensuring each iteration is increasingly aligned with the tester's expectations and the realities of software testing.

With the refined faulty code at hand, an automated integration process seamlessly incorporates the generated fault into the designated area of the application's codebase. This phase is designed to trigger an in-depth evaluation of the software's response to the introduced fault, shedding light on potential vulnerabilities and testing the robustness of existing error-handling strategies. The seamless nature of this integration, coupled with the subsequent testing, is crucial for assessing the software system's resilience, offering testers concrete insights to inform further software enhancements.

\subsection{Running Example}
To provide a clearer illustration of the methodology's application, we describe a detailed example that incorporates an input function and an example of RLHF integration.

Imagine a scenario where the tester is focused on enhancing the error handling of a specific function within their e-commerce application. The function in question, \texttt{process\_transaction}, is critical for completing customer purchases. The tester's goal is to simulate a fault where this function fails due to a timeout during a database transaction, potentially causing an unhandled exception that could disrupt the user experience.

The tester provides the following natural language description: ``\textit{Simulate a scenario where a database transaction fails due to a timeout, causing an unhandled exception within the process\_transaction function.}", alongside the current \texttt{process\_transaction} function code:

\begin{lstlisting}[language=Python, basicstyle=\scriptsize, backgroundcolor = \color{antiquewhite},frame=single]
def process_transaction(transaction_details):
    # Placeholder for database transaction code
    pass
\end{lstlisting}

This input sets the stage to generate a tailored faulty code snippet that reflects the specified scenario, integrating directly into the \texttt{process\_transaction} function. 

Based on the provided description and the context of the \texttt{process\_transaction} function, which are parsed and structured into a detailed fault specification, the LLM generates the following faulty code snippet:

\begin{lstlisting}[language=Python, basicstyle=\scriptsize,backgroundcolor = \color{antiquewhite},frame=single]
def process_transaction(transaction_details):
    try:
        # Simulated database transaction code
        raise TimeoutError(``Database transaction timeout'')
    except TimeoutError as e:
        # The fault: An unhandled exception occurs here
        print(``Transaction failed:'', e)
        # Missing exception handling logic
\end{lstlisting}

After the code generation step, the methodology presents the generated faulty code to the tester for review. The tester notices that while the generated exception simulates the desired timeout scenario, it lacks specific handling logic that would make the test more reflective of potential real-world error recovery strategies.
Hence, it inputs the following NL description: ``i\textit{ntroduce a retry mechanism instead of just logging the error}".
Using this feedback, the LLM adjusts its output learning from the tester's critique to refine the fault generation process. In response, it might generate a more sophisticated fault simulation in the next iteration:

\begin{lstlisting}[language=Python, basicstyle=\scriptsize,backgroundcolor = \color{antiquewhite},frame=single]
def process_transaction(transaction_details):
    try:
        # Simulated database transaction code
        raise TimeoutError(``Database transaction timeout'')
    except TimeoutError as e:
        print(``Attempting to retry transaction'')
        retry_transaction(transaction_details)
        # Logic for retrying the transaction upon timeout
\end{lstlisting}

This revised code snippet not only simulates the fault but also tests the application's resilience by introducing a retry mechanism, aligning more closely with the tester's expectations for comprehensive error handling.

\subsection{Implementation}

In the heart of our methodology for software fault injection lies a carefully orchestrated ensemble of components, each designed to fulfill a unique role within the process. These components work in concert to translate testers' inputs into actionable fault simulations, refine these simulations based on feedback, and integrate the results into the software testing lifecycle. In the following, we provide an in-depth exploration of the key components that underpin our solution.

\subsubsection{The Natural Language Processing (NLP) Engine}
The NLP Engine serves as the initial touchpoint for the dual-input strategy, acting as a sophisticated conduit that not only translates but also interprets the testers' inputs.
It employs advanced parsing and semantic analysis techniques, such as dependency parsing and named entity recognition, to dissect the tester's fault description. For instance, it identifies key components (e.g., ``database service'' and ``timeout'' in the NL description used for the running example) ensuring the LLM understands the context and technical requirements of the fault.
Therefore, when a tester submits a fault description in natural language alongside a snippet or file of the target code, the NLP engine undertakes a comprehensive analysis of both forms of input through advanced parsing and semantic analysis. Simultaneously, it analyzes the provided code to understand its structure, dependencies, and operational logic. This dual analysis ensures that the LLM receives a detailed, integrated input that encapsulates both the fault's conceptual framework and its practical implementation context.

\subsubsection{The Large Language Model}
Central to our methodology is the LLM, a behemoth of computational linguistics and machine learning, trained on an extensive corpus of software code and fault scenarios. This model is the engine of creativity within our solution, responsible for generating the faulty code snippets that simulate the specified software faults. But the LLM's role extends beyond mere generation; it is a learner, continuously evolving through the feedback loop established by RLHF. With each iteration, the LLM fine-tunes its understanding of software faults, growing ever more adept at producing simulations that not only meet but exceed the testers' testing requirements.

\subsubsection{RLHF Mechanism}
The RLHF Mechanism stands as a testament to the iterative nature of our methodology. Following the initial generation of faulty code by the LLM, this mechanism invites testers to engage directly with the output, providing feedback on the accuracy, relevance, and overall quality of the simulated faults. This feedback is then fed back into the LLM for code refinement. The RLHF Mechanism embodies the methodology's adaptive spirit, ensuring that the solution remains responsive to the needs of testers and the demands of the ever-evolving software landscape.

\subsubsection{Automated Integration and Testing Tool}
Completing the cycle is the Automated Integration and Testing Tool, which ensures the smooth transition of generated faults from concept to application. The tool automates the process of integrating the LLM-generated faulty code into the target software's codebase, meticulously placing each fault in its designated context for maximum impact during testing. Beyond integration, the tool facilitates a comprehensive suite of tests designed to activate the faults and observe the software's response, providing testers with invaluable insights into the system's resilience and areas for improvement.

\section{Threats to Validity}
\label{sec:threats}
\subsubsection{Data and Training Challenges}
A critical component of our methodology involves the creation of a high-quality dataset to fine-tune the LLMs for their specialized role in software fault injection. To address this issue, we leverage an existing software fault-injection tool tailored for Python applications~\cite{cotroneo2020profipy}. We adopted the tool to generate a dataset encompassing a wide array of fault scenarios across different Python software systems. Indeed, the tool allows us to systematically introduce faults into codebases and then document both the fault conditions and the resultant code changes. The collected data encompasses a diverse set of fault types, including logic errors, race conditions, memory leaks, and buffer overflows, providing a rich foundation for fine-tuning the LLMs.
The process of generating this dataset not only ensures that the LLMs have access to real-world, applicable examples of software faults but also facilitates the refinement of the models to accurately interpret natural language fault descriptions and generate corresponding faulty code snippets. 
Moreover, the ability of the SFI tool to generate this data on-demand eliminates the traditional bottleneck of data scarcity and quality, significantly accelerating the LLM training process. This ensures that the LLMs can continuously learn and adapt to new fault scenarios, enhancing their accuracy and reliability in fault simulation over time. Finally, to ensure a diverse and realistic dataset, we aim to incorporate fault scenarios from open-source projects, documented incidents, and expert simulations. 

\subsubsection{Deployment and Testing}
A pivotal aspect of deploying our advanced software fault injection methodology involves its integration with automated deployment and testing tools. This integration is crucial for showcasing the methodology's feasibility in real-world software development environments. Our tool, which is able to fully automate the generation of faulty code and its subsequent injection into the target system for testing purposes, ensures a smooth, automated transition from fault generation to system evaluation.
The ability to automate these processes significantly reduces the manual labor traditionally associated with software testing. It enables a more efficient and error-free deployment of test scenarios, directly addressing concerns regarding the scalability and practical applicability of our methodology. By leveraging this automation, we can ensure that our methodology fits within existing software development workflows without needing significant alterations or additional resources. 

\subsubsection{Efficiency and Coverage}
By automating the generation of a wide variety of realistic fault scenarios based on testers' natural language inputs, our methodology enables a more comprehensive evaluation of software resilience. This allows for the testing of software systems against a broader spectrum of potential faults, many of which may not be adequately covered by traditional testing methods.
Furthermore, the iterative refinement process facilitated by RLHF ensures that the generated faults become increasingly aligned with real-world conditions over time. This continuous improvement in fault relevance and accuracy directly contributes to more effective testing, as software systems are challenged against scenarios closely mirroring actual operational risks.

\section{Conclusion}
\label{sec:conclusion}
In this work, we proposed a methodology that adopts the capabilities of LLMs enhanced by RLHF to revolutionize the traditional practices of software fault injection. By automating the generation of simulated software faults through testers' natural language inputs, our methodology aims to simplify the fault injection process and ensure the creation of technically accurate faults and highly relevant to real-world scenarios. 

Looking ahead, we aim to validate our methodology by incorporating case studies and real-world applications. Additionally, a comparative analysis with conventional fault injection techniques will be conducted to quantitatively and qualitatively assess the improvements in efficiency, effectiveness, and coverage of fault injection.

\section*{Acknowledgment}
This work has been partially supported by the GENIO project (CUP B69J23005770005) funded by MIMIT and by the SERENA-IIoT project funded by MUR (Ministero dell’Università e della Ricerca) and European Union (Next Generation EU) under the PRIN 2022 program (project code 2022CN4EBH).

\IEEEtriggeratref{13}
\bibliographystyle{IEEEtran}
\bibliography{mybibfile}

\end{document}